\documentclass[]{PoS}
\usepackage{journals, amsmath}
\usepackage{cite}
\usepackage[inline]{enumitem}
\usepackage{siunitx}
\usepackage{booktabs,multirow,graphicx,bigdelim}

\graphicspath{{./Fig/}{./}} 

\usepackage{xspace}
\newcommand{\IceCube}{\textsc{IceCube}\xspace}

\usepackage[x11names]{xcolor}
\definecolor{myOrange}{rgb}{1,0.5,0}
\definecolor{myYellow}{rgb}{0.9,0.7,0}
\definecolor{myCol00}{rgb}{0.8,0.0,0.0} 
\definecolor{myCol01}{rgb}{0.0,0.7,0.5} 
\definecolor{myCol02}{rgb}{0.0,0.4,0.5} 
\definecolor{myCol03}{rgb}{0.2,0.6,0.4} 
\definecolor{myCol04}{rgb}{0.6,0.0,0.4} 

\usepackage{placeins}
\usepackage{changes}
\definechangesauthor[name={P G}, color=myCol04]{Pg}
\definechangesauthor[name={D F}, color=myCol01]{Df}
\setremarkmarkup{(#2)}

\title{No guaranteed neutrino  astronomy without (enough) double bang tau and downward HESE muon tracks: An updated version}

\ShortTitle{No  \protect{$\nu$}  Astronomy without enough  taus and  downward muons }

\author{\speaker{D.~Fargion},$^{a,b,c}$ {{P.~G.} {De~Sanctis~Lucentini}},$^{~d}$ M.~Yu. Khlopov$,^{~e,f,g}$ P.~Oliva,$^{h,i,c}$ F.~La~Monaca $^{a}$, P.~Paggi $^{a}$\\
~\\
\llap{$^a$} Physics Department Rome University 1, P.le A. Moro 2, 00185, Rome, Italy\\
\llap{$^b$} INFN Rome1, Rome University 1, P.le A. Moro 2, 00185, Rome, Italy\\
\llap{$^c$} MIFP, Via Appia Nuova 31, 00040 Marino (Rome), Italy\\
\llap{$^d$} Physics Department, Gubkin Russian State University (National Research
      University),\\
      65 Leninsky Prospekt, Moscow, 119991, Russian Federation\\
\llap{$^e$} Institute of Physics, Southern Federal University Stachki 194, Rostov on Don 344090, Russia\\
\llap{$^f$} APC laboratory 10, rue Alice Domon et Leonie Duquet 75205, Paris Cedex 13, France\\
\llap{$^g$} National Research Nuclear University "MEPHI" (Moscow State Engineering Physics Institute), 31 Kashirskoe chaussee, Moscow 115409, Russia\\
\llap{$^h$} Niccol\`o Cusano University, Via Don Carlo Gnocchi 3, 00166 Rome, Italy\\
\llap{$^i$} Department of Sciences, University Roma Tre, Via Vasca Navale 84, 00146 Rome, Italy\\

E-mail: \email{daniele.fargion@uniuniroma1.it}, \email{pietro.oliva@unicusano.it}, \email{desanctislucentini.pg@gubkin.ru},
\email{khlopov@apc.in2p3.fr}
}

 \FullConference{Frontier Research in Astrophysics III, May 2018
}

\abstract{\IceCube Neutrino Astronomy is considered. The $\tau$ neutrino flavor paucity and the asymmetry for the tracks suggest a dominant atmospheric charm noise. 
    The correlated cascades and tracks asymmetry with relevant statistics enforce the charm noise dominance in the data. The charm signal may explain at once the absence of correlation for the tracks data with the galactic plane and with known brightest gamma sources.  

} 

\begin{document}

%

%

\section{Introduction: \IceCube Astronomy or a new atmospheric noise?}



The high energy neutrinos up to TeV are mostly atmospheric ruled by an over-abundance of muon neutrinos tracks respect to the cascade showers that are essentially originated by atmospheric electron neutrinos and neutral current events. The ratio of such $\nu^{\text{atm}}_{\mu}$ over $\nu^{\text{atm}}_{e}$ signal is nearly $20:1$ \cite{Fargion:2013rma}.
   By atmospheric neutrinos we mean the secondary ones from the Cosmic Ray interactions with the atmosphere. Since the Cosmic Rays are charged and bent by the Galactic Magnetic Field, they and their secondary $\nu^{\text{atm}}$, are smeared and do not allow any astronomy.
   
The data collected by \IceCube from 2013 shows over 30-60 TeV a remarkable change in the relative abundance of neutrino flavor: 
indeed, in 2013-2014 a few unexpected (within atmospheric model) PeV neutrino cascade events have been first discovered \cite{PhysRevLett.111.021103}. 
Later on, more  data at hundred TeVs  confirmed  the sudden change from a dominant atmospheric muon neutrino flavor at TeVs energy,  to a more dominant cascade neutrino signal with a ratio $1:3$. These events are just spherical cascades, above tens TeV up to PeV \cite{PhysRevLett.113.101101}. Such neutrino cascade signals might be  made  by $\nu_e$,${\nu}_{\tau}$ and ${\nu}_{NC}$ interactions. Hereinafter, by ${\nu}_{\text{NC}}$ we mean the Neutral Current events made by all of the three flavors.
   The consequent interpretation for such a flavor revolution \cite{Fargion:2013rma} has been the first sign of the discovery of the UHE  Neutrino Astronomy for many.
   Most authors and  articles on \IceCube in last 7 years dealt with this newborn neutrino astronomy. However, no correlation with nearest or remarkable known (Radio,X, $\gamma$) sources and \IceCube events arose for the first years, 2010-2017. 
     The poor shower (or cascade) directionality ($\pm \ang{15}$) has been forcing us toward a more meaningful and worthful track neutrino astronomy, even made mainly by through-going or crossing induced neutrino muons \cite{FARGION2014213}. 
     
     Within first several dozens of highest energy \IceCube events (2013-2017) and with several hundreds UHECR by Auger and by TA detectors,  we tried a first  signal correlation among these  sources, with little or preliminary success  \cite{Fargion:2014jma, FARGION2017195}. More recent analogous attempts have been recently considered  \cite{barbano2019search}.
         
   Finally, in September 2017 a rare but still unique UHE through-going track event, IceCube-170922A,  has been succesfully correlated  with a previous high energy gamma Blazar TXC 0506+056, detected and active months earlier as a bright  gamma source \cite{147}. This event convinced most authors  of the \IceCube Neutrino Astronomy nature.
   
    However, there are several unsolved puzzles to be explained by such a New Astronomy.
    
    {\em Glashow resonance absence.} 
Anti neutrino UHE signals above TeVs exhibit events tracks and cascades, with similar but a little smaller cross sections, in analogy to neutrino ones in \IceCube.
A very peculiar resonant signal by an Ultra High Energy (UHE)  anti-neutrino electron at 6.3 PeV energy, might rise with an  amplified peaked  probability, as soon as it is hitting  on electron in \IceCube. These resonant event are making PeV cascade signals. Such an event, called Glashow resonance  \cite{glashow}, has not been yet discovered in a contained  High Energy Starting Events (HESE) inside \IceCube. This absence may be a new additional puzzle to be explained,\cite{SAHU20181}. Nevertheless, one may note that such a partially contained Glashow resonance event has been very recently (ICRC 2017) claimed. The hard astrophysical spectra by exponent $-2$ or $-2.2$ for UHE neutrino would be more in disagreement with such paucity. On the contrary 
  a softer charmed atmospheric spectra, with exponent $-2.7$ or $-2.9$  will be more consistent with the non observing  HESE  Glashow resonance.

The absence, at high level, of any GRB with several UHE neutrino tracks  correlation, the absence of any brightest AGN gamma sources flare in \IceCube neutrino tracks map, the absence of any Galactic plane-\IceCube anisotropy, the absence of any self UHE neutrino clustering in narrow solid angles, all of them  require an explanation.
      
There is anyway, above the mentioned IceCube-170922A correlated  with  Blazar TXC 0506+056, also a very recent, but premature, proposal for a correlation between a blazar NGC1068 source and \IceCube muon neutrino clustering at TeVs energy \cite{Aartsen:2019fau}.

But most or all, the highest UHE \IceCube alarm event among the last few years remained uncorrelated with any known  active $\gamma$ source.
For instance the brightest AGN  flare (as the huge gamma photon rain \cite{article}, on 3C 279 on June  2015) or highest energetic GRB  do not find any \IceCube UHE neutrino precursor or correlated track partner.
    
The most recent puzzle is also the absence (or paucity)  of any clear tau neutrino event among several dozens (28-34) of UHE neutrino cascades above hundred or 90 TeV. Very recently, since June 2018, \IceCube offered and claimed two tau candidates \cite{Juliana2018Poster174,Stachurska:2019wfb}. 

In our opinion, as shown below, both are probably no tau signal.
It is also to be underlined that among (28-34) cascades, within our alternative atmospheric charm neutrino scenario, there is anyway room for at least one charmed tau neutrino\cite{Fargion:2018yoqV4}.
\IceCube  explanation is a suppressed  tau neutrino  detection, as shown in Fig.\ref{Fig16} and discussed below.
In addition to the above absence there is  also the \IceCube average neutrino flavor well consistent with a charm atmospheric one. See Fig.\ref{Fig12}.
             
Associated to the tau neutrino puzzle there is the surprising paucity of downward vertical UHE muon neutrino tracks HESE or trough-going above hundred TeV energy. Their absence, even respect the upward ones, may be indebt to the \IceCube Top veto for eventual correlated downward airshower. Consequently, this downward paucity from the South sky, not observed in the upward North one,  might be explained by an  atmospheric neutrino noise, consistent with  most of the mentioned astronomical correlation absence.

\subsection{Possible  \IceCube astrophysical hidden origin }            
 To allow a reasonable escape road for the \IceCube astrophysical nature, to make it consistent with missing correlation,  we offer  a  very old but peculiar cosmological model that proposed two galactic populations. One population is near at a few redshift while the second is located at far tens redshift and is somehow able to hide the photon signals by distance, dust and opacity. 
 The hadronic interaction  secondaries, source  of both gamma and neutrinos, are reaching us in different rates: the coeval UHE gamma signals  are shield from us respect to the more transparent and penetrating \IceCube events.
 Therefore, neutrinos from large redshift (z> 10) are no longer correlated with far but hidden photon signals.
 These two clustering epochs are due to the multi-fluid clustering in cosmic expansion \cite{Fargion1983}. 
 This two ages for galactic formation may be related, for instance,  to the cosmic helium and hydrogen gas different recombination and to different temperatures while cooling, leading to delayed clusterings and to different galaxy formation times.
 
 More exotic possibilities for different clustering epochs might be also related to the eventual cold dark matter  component (as neutrino, neutralino), either by their light or heavy masses, by their different interactions or by different decouplings \cite{Antonelli1981}, leading to mixed  baryon and dark matter multi-fluid clustering \cite{Fargion1981,Fargion1983,Khlopov_2006}.

\subsection{Alternative scenario: a dominant charm noise}

The two above mentioned galactic populations, are  somehow in disagreement with the unique event IceCube-170922A  observed  at z= 0.3.
A more convincing and testable approach is offered in the present article. It is based on a noisy charm atmospheric component smearing \IceCube data and hiding the minor astrophysical signals.
Indeed, the noted flavor sudden change might be mimicked also by the simplest rise of the charmed atmospheric component. This noise arises exactly at the same hundred TeV energy windows where \IceCube revealed the UHE events.
Many authors agreed that such a charmed atmospheric neutrino component must exist and plays a role\cite{Vissani2019}, but most of them considered the charm signals as secondary component, of one third or less of the present observed flux, in the \IceCube rate. Therefore, most authors, as \IceCube researchers themselves,  remained convinced of the main UHE neutrino astrophysical nature.
 We remind anyway that the charmed atmospheric spectra derived by  Cosmic Rays, CR, models is not well defined. It is known within a factor two. Moreover, it is dependent on the real, but unknown, PeVs energy CR composition, leaving much room for its eventual underestimated relevance. 
 Hence, we are still offering the present prompt charm atmospheric noise scenario for the \IceCube data interpretation\cite{Fargion:2018yoqV4}.
 
%

\section{\IceCube sample data: 2018-2019}    

%
%
\begin{figure}[!t]
\begin{center}
\includegraphics[width=0.7\textwidth]{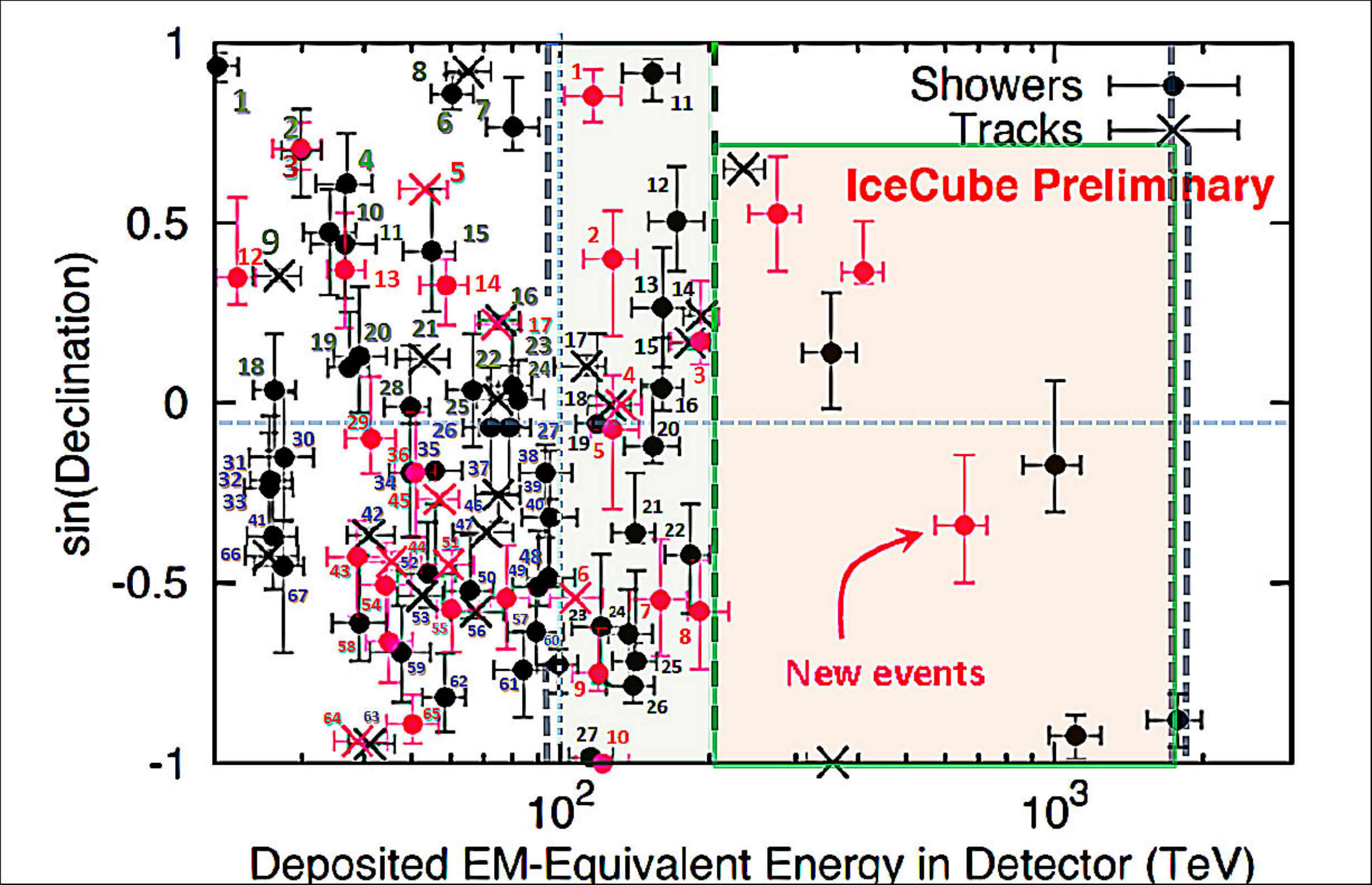}
\caption{The last (June 2018) 103 HESE data set\cite{Wandkowsky2018Poster175,Taboada2018Talk} On the vertical axis the HESE events arrival direction, in sin(Declination) within celestial coordinate, while on the horizontal one the deposited EM-equivalent energy released in the \IceCube detector. The energy lower bound threshold has been reduced to 100~TeV, see dotted line (or even at 90~TeV as in the colored areas); the two tau candidate events to inspect are just among 27 events above 100 TeV and  9 above 200 TeV, for a total of 36 events. Adding the events in the range 90-100~TeV the total number of events rises to 42. Among them we may observe 34 shower that might be made not only by an electron neutrinos and more rarely by a neutral current, but also by tau neutrinos, on average nearly a dozen.  On the extreme sides of the dectection range: the highest energetic shower at 2 PeV and a weak event at 89 TeV are the two Tau shower candidates  offered  by \IceCube. 
}\label{Fig4B}
\end{center}
\end{figure}
%

     \IceCube data has not been published in detail for several years.
    Only maps  and energy-coordinate spectra have been released in conferences in 2018-2019.
 We will use the 103 HESE event data of 2018 \cite{Taboada2018Talk, Wandkowsky2018Poster175}, see Fig.\ref{Fig4B}, and the most updated UHE neutrino map \cite{barbano2019search} of 2019, see Fig.\ref{FigXX}, where we also combined the early information on the UHE HESE \IceCube neutrino events shown in \cite{Taboada2018Talk, Wandkowsky2018Poster175}, to disentangle external tracks (the through-going) from the internal HESE ones, see Fig.\ref{Fig4B} and Tab.\ref{tab:table1}.

We can distinguish events of three types: first the High Energy Starting Events, or HESE, tracks originated inside the \IceCube detection volume, second the HESE cascades originated and completely contained in the same volume, and third the tracks not HESE, but through-going, originated outside and crossing the whole detection volume. We can count $26$ HESE tracks, $76$ HESE cascades, for an overall $102$ HESE events and nearly 60 through-going tracks shown in Fig.\ref{Fig4B}.
The total track number, HESE and external, is $86$ as shown in Fig.\ref{FigXX}.

Even if they were originally presented togheter \cite{barbano2019search}, we distinguish them via the colour label, as shown in Fig.\ref{FigXX} and explained in the relative caption. 

We also maked in the Fig.\ref{Fig4B} the energy ranges associated to the two tau candidate event claimed by \IceCube at 89TeV and 2 PeV \cite{Juliana2018Poster174,Stachurska:2019wfb}.

\begin{figure}[!t]
  \begin{center}
    \includegraphics[width=0.65\textwidth]{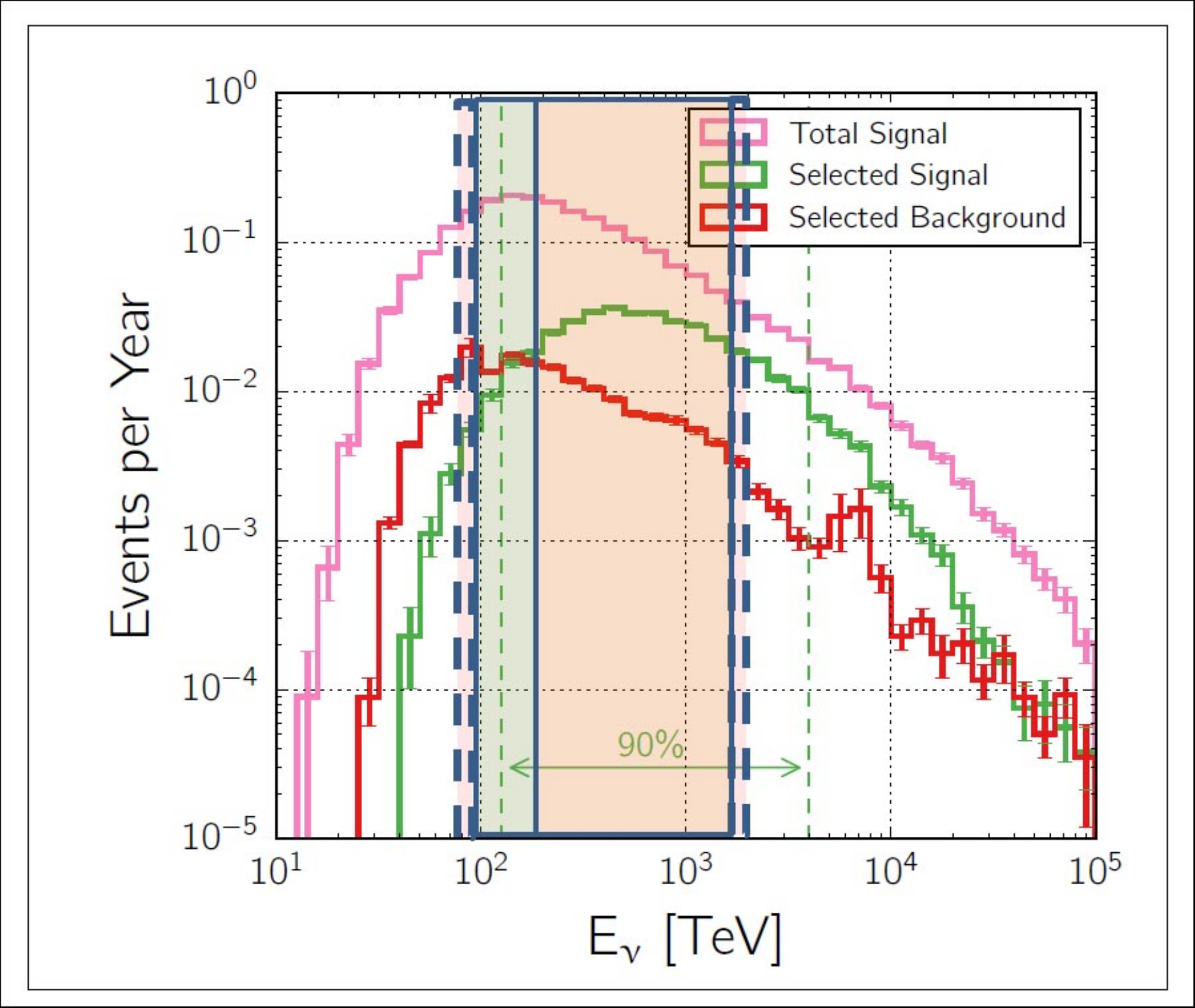}
    \caption{The \IceCube expected event rate of Tau neutrino as a function of the energy for signals  (green curve) and noises (red curve).
    Note that the lowest energy tau event, at 89 TeV, is  the most un-probable signal. Noise is four times larger than signal.
    Note also that the highest Tau event at 2 PeV it was revealed on 2014. It took four year to identify it as a double bang. Incidentally a 2 PeV tau would track mainly about 100 meters, while the double signal observed  is five times smaller, it is just at 16 meter of separation. In (opposite)  analogy a lowest 89~TeV signal would track mostly around 4.5 meters, while the observed double bang for the weakest event is this time four times longer, around 17 meters. The most surprising signature, statistically more unexpected is the location of both of the event in worst and most un-probable curve regions. The colored rectangular areas (on the left, the most crowded one, on the right, the most probable one) are surprisingly empty of any tau events. See the discussion in the text. }
    \label{Fig16}
  \end{center}
\end{figure}

The correlated Fig.\ref{Fig16} shows the corresponding \IceCube energy ranges in function of the detection ability and probability \cite{Usner2018SearchPhdThesis}.


It is easy to underline the low probability  to be both of them  real tau signal events. For instance at lowest energy, at $(80+9)$ TeV one,  the expected atmospheric neutrino noises exceeds by a factor four any astrophysical signal \cite{Usner2018SearchPhdThesis}. The consequent tau track at $90$ TeV would be expected at around 4.5 meters, and not at the 17 observed meters.  

It is also hard to believe that the largest energy event, the so called Big Bird at 2 PeV event discovered on 2014, has been considered a true double bang signal after five years of analysis. 
Indeed, the observed first bang energy released by the 2 PeV event is much larger respect to the second one. 
The theoretical tau double bang should show an opposite energy imprint: first a small and later a larger second shower, twice times or more brighter \cite{Learned:1994wg}. 
Moreover, because $L_{\tau}=49 (E_{\tau}/{PeV})$ meters, the tau distance at 2 PeVs should range around hundred meters and not just the 16 meters observed.

%
%

\subsection{Tau neutrino as a probe of Neutrino Astronomy}
As we outline in the title, to be convinced of the 
 UHE astrophysical signals nature one would like to observe the ${\nu}_{\tau}$ presence in \IceCube UHE events. Its presence as abundant as the other flavours ${\nu}_{\mu}$ and ${\nu}_{e}$  will  clearly confirm the \IceCube astrophysical origin. This full flavor mixing is guaranteed even by the tiny neutrino masses  and by their splitting and mixing at the atmospheric and solar observed levels.
               
Indeed, the astrophysical neutrinos are mostly oscillated and mixed along the widest stellar and cosmic flights, leading in most scenarios to a complete averaged equal three flavor ${\nu}_{\tau}$, ${\nu}_{\mu}$, ${\nu}_{e}$ probability. Therefore, ${\nu}_{\tau}$ might and should be present and possibly observable in the several \IceCube events above hundred TeV.
 This is not yet (well) observed.
        
The UHE muon neutrino ${\nu}_{\mu}$ are easely produced by pion and Kaon decay in our atmosphere up to TeVs. Instead, their secondary energetic ${\nu}_{e}$ are hardly produced by the muon decay in-flight.
Their ${\nu}_{\tau}$ component is absent, even by the ${\nu}_{\mu}$  oscillation along the Earth.
Moreover, the UHE ${\nu}_{\tau}$  are  negligible even in the charmed atmospheric component because they are suppressed by a smaller cross-section for the charmed atmospheric ${\nu}_{\tau}$ component \cite{Enberg:2008te}. 
 
Therefore, an evident tau presence 
 will mark the astrophysical origin of the \IceCube events. 
 This explains the importance of the tau role, a key role, and clarifies our persistence in seeking the definitive discovery of the tau neutrino\cite{Fargion_2002,Fargion_2004}.

\section{Tau neutrino detection}  
How could UHE  ${\nu}_{\tau}$ be detected in \IceCube?
    
    
A UHE $\nu_e$ with energy in the range from several TeVs up to PeV, traveling through the detector volume, interacts with the nuclei creating a shower or cascade: several meters of tree rich ramification in pions and electromagnetic secondaries; 
the secondaries Cherenkov optical photons of this cascade will be diffused in ice by random walk into a spherical shape. For a shower of hundreds TeV the size of the sphere can reach up to hundreds meters.

Largest PeV cascade events may extend their Cherenkov lightening in several hundreds meter spherical radius, filling most of the detector volume.

By the timing spread or diffusion in \IceCube it is possible to estimate somehow the neutrino directionality within a wide solid angle ($\sim \pi\theta^2$ with $\theta=$ \SIrange[range-units = brackets]{15}{30}{\degree} ). 
In general, also the Neutral Current and the relevant tau by ${\nu}_{\tau}$ event at tens TeV  may lead to a similar cascade as the $\nu_e$ one. Indeed, fast tau decay overlaps its birth place hiding the short track inside the light \textit{bang}.
There are observable different signatures, as discussed soon, for more energetic ${\nu}_{\tau}$ events above hundred TeV. Therefore, in \IceCube present resolution, UHE neutrino cascades up to nearly hundred TeVs are  well hiding  their own neutrino flavor identity.  
However, above the hundred TeV threshold the $\tau$ fast decay ($2.9 \cdot 10^{-13}$ sec.) may be separated and resolved by their ${\nu}_{\tau}$ birth cascade, both in distance and in time. 
This occurs because the $\tau$ relativistic boosted life is leading to two separated cascades, two \textit{bangs}, linked by $49$  $\cdot (E_{\tau}/PeV)$ meters track size \cite{Learned:1994wg}:  a first UHE ${\nu}_{\tau}$ hadron nuclear interaction  with a fourth or a third of the primary ${\nu}_{\tau}$ energy; then a later $\tau$ decay with the rest major part of the ${\nu}_{\tau}$ energy. These two bangs may be both contained in the present km$^3$ detector volume. 
Because of the actual detector size and  the optical array resolution, the best detection energy window for such $\tau$ double bang is with tracks from tens to hundreds meters: an energy within a range of nearly 100 TeV up to a few or several PeV,  as shown in The Fig.\ref{Fig16}. We note the maximal probability to detect such a double bang in \IceCube \cite{Usner2018SearchPhdThesis}, occurred  around $400$ TeV, quite distant both from the $89$ TeV and 2PeV observed events at the extreme edges shown in Fig.\ref{Fig16}.
  


\subsection{Tau neutrino airshower}
 An analogous more filtered tau neutrino signal would be an UHE PeV-EeV up-going tau escaping from  a mountain or the Earth airshowering to the sky. In synthesis the double bang \cite{Learned:1994wg} considered above in \IceCube occurs with a first bang  inside the mountain or inside the Earth crust, and the second bang in the air. This consequent tau airshower Astronomy has been offered since nearly 20 years  \cite{Fargion:1999se,Fargion_2002}. Unlike isolated upward muon tracks, the upward tau airshower expands in a vast area (km square size) its presence, by a million or even thousand of billion secondary traces. The event is an ideal filter and amplifier of Neutrino Astronomy.   
 Tau airshower has been studied in more detail in flight across the Earth opacity \cite{Fargion_2004}. It became more and  more experimented  only in the recent decade. 
The upward tau airshower have not yet been observed. This tau airshower signal, often referred by an improper name as a skimming neutrino \cite{Feng_2002}, could and should offer  the ideal noisy free Neutrino Astronomy. Several present and on-going  experiment are searching for such tau airshowers upgoing from the Earth toward the horizontal sky. AUGER or TA array UHECR records by fluorescence lights, by upgoing horizontal airshower, might soon reveal such EeV neutrino traces.

\subsection{\IceCube events} 

Up to early 2018 Icecube claimed the 102 HESE events, 76 cascades and 26 tracks, all above 20 TeV energy, with no $\tau$ detection. Since June 2018  two $\tau$ event candidates have been finally claimed.  However, as we shall argue, both of them are not a probable tau candidate signal because of the disagreement between their observed versus expected $\tau$ signature. Therefore, tau neutrino signals among last \IceCube HESE 76 cascades, in particular the most energetic ones above 100 TeV (28 cascades), or above 90 TeV (34 cascades), should contain nearly a dozen of $\tau$ double bang.
The only two observed events are too few respect the predictions: the lowest energy one is located in a range of energies where is polluted by too much noise and the second is distributing the energy between the two showers in a way opposed to the expectations.

\subsection{The flavor mixing weight in \IceCube}  

%
\begin{figure}[!h]
\begin{center}
    \includegraphics[width=0.7\textwidth]{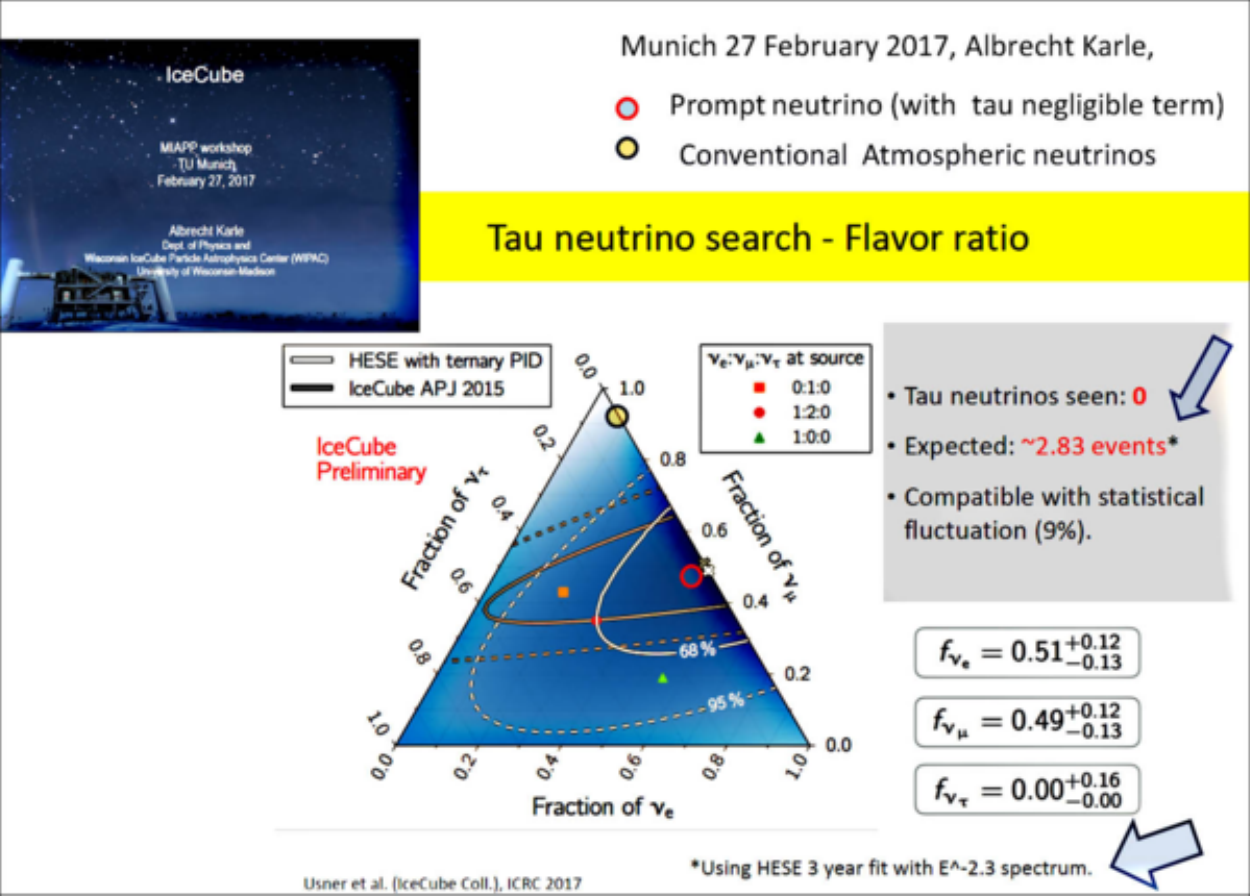}
\end{center}
\caption{Elaboration of Fig.\ref{FigXX} from \cite{usner2018search}. A recent article based on 2017 record by \IceCube  has been favoring not the common expected astrophysical flavor ratio $1:1:1$ but a flavor combination tuned with the charmed case:~$\phi_{\nu_{\mu}}$=$\phi_{\nu_{e}}$. More exactely:~$\phi_{\nu_{\mu}} = 0.49\%$ and $\phi_{\nu_{e}}= 0.51\%$. The blue ring on the top shows the  muon atmospheric  dominated role with a tiny  $5\%$ component of the electron ones. The presence of a small tau component  also at $5\%$ ratio,  occurs even for the atmospheric charmed neutrino component, as for the case of one or two tau  labeled by a red ring that is well correlated with the white cross of the observed data. It should be noted also the expected tau event rate of $2.83$ event every three years is well comparable with the expected 9 events discussed in our article for the whole 2011-2019 period \cite{Fargion:2018yoqV4}.
    }
    \label{Fig12}
\end{figure}
%


Without the two tau events in 2018 one may inspect the whole \IceCube neutrino map in a flavor triangle map. For instance the weighted map of 2017 by \IceCube, see Fig.\ref{Fig12}. One may observe that the most probable flavor point for an astrophysical model (red dot) is near the center by an equipartition  flavor ratio : $\nu_e:\nu_{\mu}:\nu_{\tau}= 1:1:1$. 
The most probable observed signal (dense area) is more centered on a very different flavor side: $\nu_e$:$\nu_{\mu}$:$\nu_{\tau}$= $1:$$1:$$0$. This is just the expected charmed atmospheric noise that we are here defending. The tiny component of an eventual charmed tau component will lead to a similar ratio: $\nu_e$:$\nu_{\mu}$:$\nu_{\tau}$= $1:$$1:$$0.05$, as shown by a red ring in the  Fig.\ref{Fig12}. 
Naturally the new two tau events in 2018 may partially correct the flavor weight in favor of an astrophysical component. Nevertheless, the two mentioned tau events are not both ideal convincing signals.
  
Moreover,  one may observed by that the same \IceCube in 2017 presentation, see Fig.\ref{Fig12}  was foreseen a tau rate event of nearly three  event every three years, leading to almost nine event today. In agreement with our first estimate,  in disagreement with the too few and improbable two tau \IceCube events.

\section{Upward-Downward  HESE events and tracks in \IceCube }
\begin{figure}[!hb]
\begin{center}
    \includegraphics[width=0.8\textwidth]{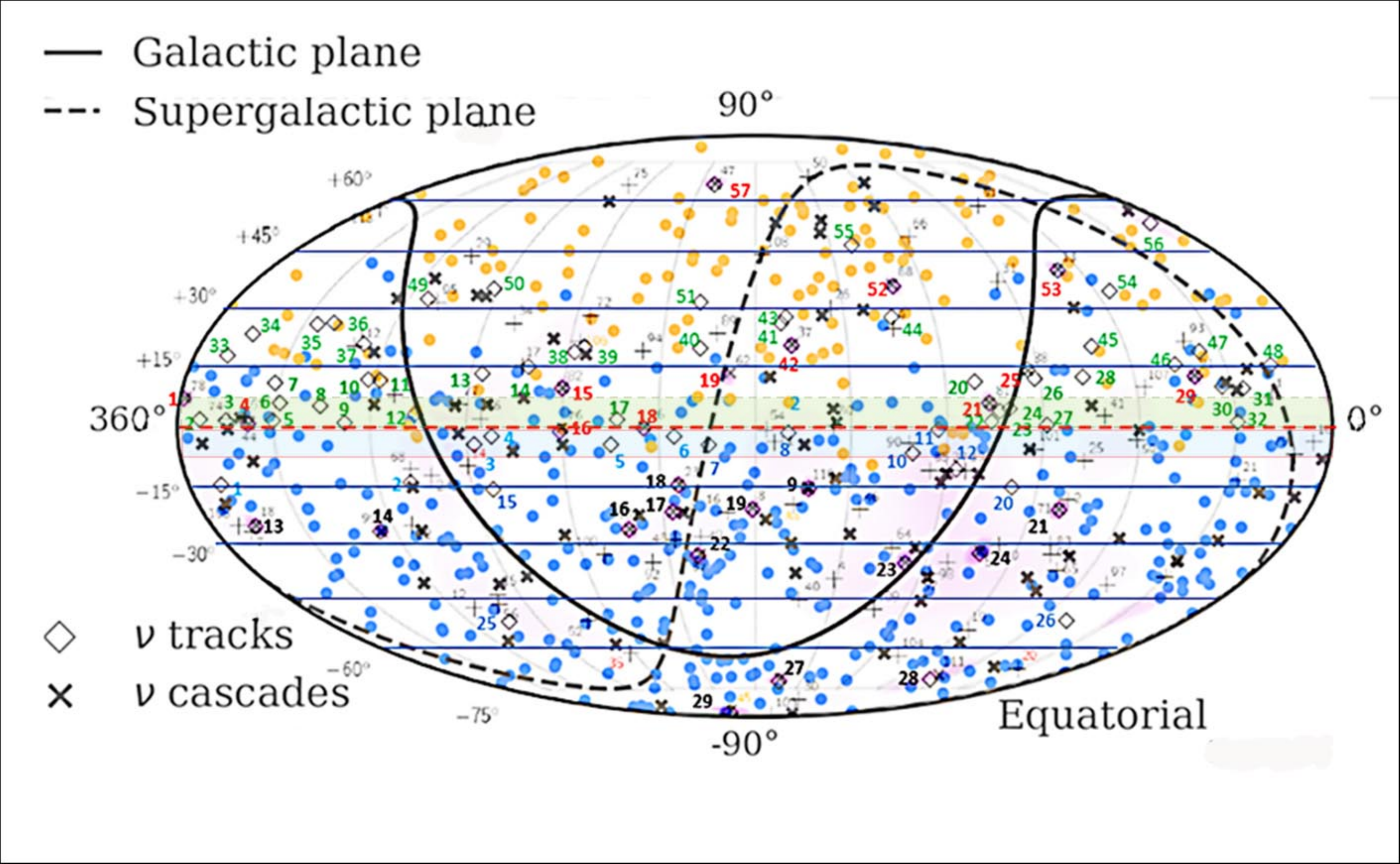}
\end{center}
\caption{The most recent \IceCube event map, combined with several collaborations shown in ICRC 2019 \cite{barbano2019search}. ANTARES contributed  by 3 events. AUGER and TA (Telescope Array) UHECR events are labeled respectively by blue and orange disks. The $\nu$ events are both originated by $\nu_{\mu}$ leading to tracks (a diamond symbol) or a   $\nu$ cascades, labeled by a dark cross, made possible by either a $\nu_{e}$, $\nu_{NC}$ or by $\nu_{\tau}$} UHE neutrino interaction. The $\nu_{\mu}$ tracks  themseles might be originated either inside the \IceCube  named HESE (high Energy Starting Event) or they are made by through-going muons whose  $\nu_{\mu}$ interaction and birth it is outside the IceCube $km^3$ itself. These through-going muons tracks are disentangled by the overlap of  ICRC 2019 \cite{barbano2019search} by HESE events map 2018 \cite{Wandkowsky2018Poster175}. The cascade events are much smeared in their detection. Therefore, they are counted only in widest solid angle, just to verify the upward downward ratio.
The HESE track event presence is marked by an underlined violet mini cross over (the otherwise) empty diamond. Remaining empty diamonds are just labeling through-going muon tracks born outside the $km^3$ detector.   We pointed the upward (from the North sky) tracks by numbering them in horizontal lines among sequence of 
    \SIrange[range-units = brackets]{0}{15}{\degree},
    \SIrange[range-units = brackets]{15}{30}{\degree},
    \SIrange[range-units = brackets]{30}{45}{\degree},
    \SIrange[range-units = brackets]{45}{60}{\degree} and
    \SIrange[range-units = brackets]{60}{90}{\degree} solid angles. The Upward tracks are label by the red colour (HESE) number while  green colour numbers  are the through-going muons events. In analogy the downward (from South Pole Sky) tracks are counted in
    \SIrange[range-units = brackets]{0}{-15}{\degree},
    \SIrange[range-units = brackets]{-15}{-30}{\degree},
    \SIrange[range-units = brackets]{-30}{-45}{\degree},
    \SIrange[range-units = brackets]{-45}{-60}{\degree},
    \SIrange[range-units = brackets]{-60}{-90}{\degree} 
    lines. The black numbers label the HESE contained events  while the blue numbers are pointing to through-going muons. The dashed red line describe the IceCube horizons in celestial coordinate. As in the text the whole sample is strongly anisotropic, enhanced at horizons. Moreover there is an additional asymmetry among horizontal upward (more abundant) and downward (less events) tracks. In the text the main consequences.
    \label{FigXX}
\end{figure}

A very recent August 2019 \IceCube  map shown in ICRC, is containing more useful updated information. There are hundreds of recent UHECR events (by AUGER and TA, Telescope Array), all the \IceCube cascade HESE events (76 events) and muon HESE neutrino tracks (26), reaching nearly  86 total tracks,  see Fig.\ref{FigXX}. In this figure we overlap the HESE muon tracks with 2018 map data\cite{Wandkowsky2018Poster175,barbano2019search}. This comparision permited us to disentangle trough going muon tracks (by UHE neutrino, crossing \IceCube from side to side) from HESE contained muon tracks. 
This disentanglement is shown by differnt colors for the event numbers in Fig.\ref{FigXX}.

We grouped the events in following ranges of zenith angles above and below the horizon: 
above  \SIrange[range-units = brackets]{0}{15}{\degree},
              \SIrange[range-units = brackets]{15}{30}{\degree},
              \SIrange[range-units = brackets]{30}{45}{\degree},
              \SIrange[range-units = brackets]{45}{60}{\degree} and
              \SIrange[range-units = brackets]{60}{90}{\degree};
and below                \SIrange[range-units = brackets]{0}{-15}{\degree},
              \SIrange[range-units = brackets]{-15}{-30}{\degree},
              \SIrange[range-units = brackets]{-30}{-45}{\degree},
              \SIrange[range-units = brackets]{-45}{-60}{\degree},
              \SIrange[range-units = brackets]{-60}{-90}{\degree} .             
About the upward tracks we colored in red the HESE ones and in green the trough-going ones. 
About the down-ward tracks we used the black for the HESE and the blue for the trough-going ones.

The track resolution is within $\ang{1}$ and allows us to discuss  some remarkable  angular anisotropy and asymmetry for tracks.

 We made similar count for the cascades: because of their wider and smeared angular resolution, about $\ang{15}-\ang{30}$, we classify them just for the sign of the arrival direction in the upward and and in the down-ward groups.
 
          
 We analized these events in the view of an astrophysical or an atmospheric charmed nature deriving the correponding expected rate and anisotropy and asymmetry.

\subsection{Upward-Downward  HESE cascade and tracks }

%
%

  Let us remind here that by upward event we mean a signal from North sky and by downward event a signal coming from the South sky.
  The straightforward message that arises in the cascades and tracks map is the opposite up-down ratio. There are 29 upward and 47 downward cascades. Their ratio is  $R^{\;c}_{Down/Up}=~47/29~=~1.62$.
   On the contrary, there are 57 upward tracks (HESE and trough-going) and 29 downward ones. Their ratio is $R^{\;t}_{Down/Up}=~0.5089$ ($R^{\;t}_{Up/Down}=~1.965$). 
   
   To be underlined the opposite cascade (poorer upward) and tracks (poorer downwards) ratio signature. Their ratio $R^{\;c}_{Down/Up}/R^{\;t}_{Down/Up} \geq 3.18$  has different key explanations. 
   

The upward cascade remarkable asymmetry (R= 1.62)  and its paucity might be related mainly to the Earth opacity for UHE neutrinos.

Indeed, the HESE tracks show a negligible asymmetry $R^{\;t-HESE}_{Down/Up}$= 14/13= 1.07, far from the cascade ones, R= 1.62.
It might be due to a competitive detection efficency veto by downward atmospheric noise. 

 Tracks by through-going muons, of external events, are quite long outside the detector, as we will estimate later nearly 2.5 km. The $1.45$ kilometer thick ice layer above the detector makes less probable the downward through-going muons signals compared to the upgoing UHE ones.
 Upward trough-going tracks below ($\ang{-34}$) are coming  mostly interacting  inside the rock Earth, where is the main calorimeter. The same Earth opacity for through-going UHE neutrino plays an opposite relevant role, but it is overcome by previous efficency. They are 44 upward and 15 down-ward  $R^{\;t-Trough}_{Down/Up}= 0.341$.
 
 This very strong asymmetry respect cascades is due to $\nu_\mu$ interaction probability, to their lenght in ice and rock and to the strong downward veto used to exclude the atmospheric noise: when a downward muon has a correlation in time and direction with an an atmospheric event, it can fall under the veto triggered by the Ice Top CR array. It may also rarely happen that a through-going track comes together with a twin charmed companion and finish to be discarded as atmospheric noise.

 Cascade are so much smeared in angular direction that their eventual atmospheric shower and origination avoid the IceTop veto.
   Moreover, there is no UHE electron charm patners able to come inside the \IceCube.
   Therefore, such downward cascade are in proportion more abundant than the downward tracks.
   
 This asymmetry for contained HESE events already suggested the atmospheric polluted dominant role. Just assuming a symmetric up-down ratio to satisfy the HESE tracks, with  47 downward  cascades  versus 29 upward ones, the asymmetry will be statistically quite unprobable: by a binomial probability calculations it will be below $1.1\%$. This probability is indeed mitigated for the observed ratio $R^{\;t-HESE}_{Down/Up}=14/13$: $P=2\%$.

\subsection{Horizontal Anisotropy and Asymmetry in \IceCube}
 We already mentioned that the trough-going muon neutrino tracks are \textit{long} and exceed the \IceCube size crossing the detector side to side. This \textit{de facto} increases the observable neutrino volume. The increase, being only longitudinal, grows linearly with the muon characteristic length.

 At hundreds of TeV the muon track in the ice is less than ten kilometers long due to energy losses\cite{Fargion_2002}.
 The search of selected  UHE astrophysical neutrinos  at hundreds TeV  makes them  originated mostly nearer than 10km  from the detector \cite{Fargion_2002,Fargion_2004}.

We may reach a first average estimate of the muon track outside the \IceCube assuming in a first approximation the ratio of the trough-going event respect to the HESE tracks of recent 7 years: 59 trough-going events vs 26 HESE, with a ratio of 2.26. 
Hence, we may assume that the average muon tracks outside the detector is about 2.2 Km long. 
An additional estimate may be reached by similar ratio found in the horizontal angular strip of $\ang{0}-\ang{15}$: 23 through-going events versus 9 HESE, leading to 2.55 ratio.
In conclusions the muon tracks outside  \IceCube are on average well below three kilometers long, let us assume for the following calculations $L_{ext}=2.5$ km.

The solid angle at $\pm \ang{15}$ above and below the horizon in first approximation insists  on a comparable slanth depth.
This would be true for a detector located at least 0.65 km over the ice. However, the detector sits  just at 370 meters from the rock. Therefore, a tiny component of the detector volume may be recording UHE neutrinos produced also in the rock, partially breaking the up-down symmetry. The much narrow solid angle of $\pm\ang{7.5}$  guarantees that \IceCube is observing comparable slanth depth in ice above and below the horizon.

The $\pm\ang{7.5}$ solid angle contains 25 events, as shown in Fig.\ref{FigXX} and reported in Tab.\ref{tab:table1}: \SIrange[range-units = brackets]{0}{7.5}{\degree} with 17 upward events, and \SIrange[range-units = brackets]{-7.5}{0}{\degree} with 8 downward.
        
The probability $P$ that this asymmetry up-down occurs, assuming as we said a symmetric astrophysical signal, is near to
\begin{equation}
    P= 3.22\cdot 10^{-2}.
\end{equation}
In the more statistical populated range $\pm \ang{15}$ (with a tiny almost negligible pollution rate from the the rock) the consequent probability due of the larger numbers (32 upward versus 12 downwards) is:
\begin{equation}
    P= 1.19 \cdot 10^{-3}.
\end{equation}
All these are  additional hints that there is a peculiar veto for downward tracks related to the ICE TOP detection of polluted airshowers.
This estimate may be extended to arrival  solid angle as large as $\pm \ang{30}$, with some cautions because of the Earth opacity that may however suppress, but not increase, the upward versus downward track ratio. The underground rock increases with relevance the upward rate. The two effects may not always be comparable and may not mutually wipe out.  
The upward tracks (48) versus the downward ones (21) imply a probability to occur as small as: 
        \begin{equation}
            P= 4.57 \cdot 10^{-4}.
        \end{equation}
The analogous estimate for similar upward (38) and downward (13) asimmetry for through-going tracks may occur only for:
        \begin{equation}
            P= 2.11 \cdot 10^{-4}.
        \end{equation}
These last two probabilities might be considered with much caution than the previous ones.

What stated above may strongly suggest a relevant atmospheric veto for the downward events due to their major atmospheric nature, veto absent instead for the upward events coming across the Earth from the North sky.
          
One may be tempted to suggest a dominant common atmospheric origin via Kaon and Pion to explain the overabundance of the horizontal signals. 
Indeed, for the ideal angular distribution in spherical symmetry, we may count overabundance in horizontal solid angle ranges of $\ang{0}-\ang{15}$ that insists over a solid angle fraction of the sky (over the whole $4\pi$),  $\sin (\ang{15})/2$ = 0.1294.
 The probability to find 32 tracks  inside such a narrow  upward solid angle of $\ang{0}-\ang{15}$
within 86 tracks is:
\begin{equation}
   P= 8.57 \cdot 10^{-9}.
\end{equation}
This horizontal track overabundance  cannot be indebt to largest decay distances for skimming pions and Kaons at far horizons edges: indeed $\nu_e$ may be generated by Kaons and Pions with a suppressed ratio respect to $\nu_\mu$  (of $\sim 5-10\%$) in disagreement with the superior abundance of the recorded cascades versus muon tracks.
Therefore, the charmed atmospheric role is the fitting one leading as we mentioned to cascade event number more numerous than tracks, see Fig.\ref{Fig12}.

There is also an independent tool  based on the asymmetry among upward and downward cascades and tracks above 80 TeV. As shown in Fig \ref{Fig4B} the upward HESE tracks are 6 while the upward cascades are 10. Their ratio is $R= (10/6)= 1.666$.

The downward cascades above 80 TeV are nearly 24 while the down tracks are 2.
Their ratio is $R= (24/2)= 12$,   in deep contrast with previous one $R= 1.666$.
              
Assuming as true the upgoing ratio R= 1.666, the consequent downward configuration binomial probability to occur becomes:
\begin{equation}
  P=  \frac{26!}{24!\cdot2!} \cdot \left(\frac{10}{16}\right)^{24} \cdot \left(\frac{6}{16}\right)^2  = 5.7 \cdot 10^{-4}.    
  \label{formula:cas-track}
\end{equation}
   A quite unprobable rate of events:
   too few downward muon tracks or too many downward cascades. Only an atmospheric veto may lead to such a paucity of the muon downward tracks. 
   
   To make the cascade and muon rate comparable in up and down sky we need nearly 14.4 muons versus the observed 24 cascades. The observed tracks are instead just 2. As a first preliminary estimate the possible polluted signals exceeds by a factor $7.2$ the eventual astrophysical ones.
          This may mean nearly $13.9\%$ of astrophysical signals in a charmed ruling sky.
Therefore, for us the natural explanation remains the charm atmospheric noise signal able to hide most of the Neutrino Astronomy in a smeared track sky.


\begin{table}
\centering
\resizebox{0.7\linewidth}{!}{%
\begin{tabular}{cccccccccc} 
\toprule[1.5pt]
 & \multirow{2}{*}{Ranges} & \multicolumn{4}{c}{Tracks} & & \multicolumn{2}{c}{Cascades} & \multirow{2}{*}{All} \\ 
\cline{3-6} \cline{8-9}
\multicolumn{1}{l}{} &  & \multicolumn{1}{l}{All} & \multicolumn{1}{l}{HESE} & \multicolumn{1}{l}{T-G} & \multicolumn{1}{l}{Sum} & \multicolumn{1}{l}{} & \multicolumn{1}{l}{Group} & \multicolumn{1}{l}{Sum} & \multicolumn{1}{l}{} \\ 
\midrule
\multirow{5}{*}{\begin{tabular}[c]{@{}c@{}}
\rotatebox[origin=c]{90}{\parbox[c]{2cm}{\centering Upward~or North~Sky}}
\end{tabular}} 
 & \SIrange[range-units = brackets]{60}{90}{\degree} & 1 & 1 & 0 & \rdelim\}{5}{4mm} \multirow{5}{*}{57} & & \multirow{1}{*}{\}~~~~2} & \rdelim\}{5}{4mm} \multirow{5}{*}{29} & \rdelim\}{10}{4mm} \multirow{10}{*}{76} \\
 & \SIrange[range-units = brackets]{45}{60}{\degree} & 2 & 0 & 2 & & & \rdelim\}{1}{4mm}  \multirow{2}{*}{10} & & \\
 & \SIrange[range-units = brackets]{30}{45}{\degree} & 6 & 2 & 4 & & & & \\
 & \SIrange[range-units = brackets]{15}{30}{\degree} & 16 & 1 & 15 & & & \rdelim\}{1}{4mm} \multirow{2}{*}{17} & & \\
 & \SIrange[range-units = brackets]{ 0}{15}{\degree} & 32 & 9 & 23 & & & & \\ 
%
 \cline{2-8}
\multirow{5}{*}{\begin{tabular}[c]{@{}c@{}}
\rotatebox[origin=c]{90}{\centering \parbox[c]{2cm}{\centering Downward~or South~Sky}}
\end{tabular}} 
 & \SIrange[range-units = brackets]{  0}{-15}{\degree} & 12 & 1 & 11 & \rdelim\}{5}{4mm} \multirow{5}{*}{29} 
 & & \rdelim\}{1}{4mm} \multirow{2}{*}{28} & \rdelim\}{5}{4mm} \multirow{5}{*}{47} & \\
 & \SIrange[range-units = brackets]{-15}{-30}{\degree} & 9 & 7 & 2 & & & & \\
 & \SIrange[range-units = brackets]{-30}{-45}{\degree} & 3 & 3 & 0 & & & {\rdelim\}{1}{4mm} \multirow{2}{*}{15}} & & \\
 & \SIrange[range-units = brackets]{-45}{-60}{\degree} & 2 & 0 & 2 & & & & \\
 & \SIrange[range-units = brackets]{-60}{-90}{\degree} & 3 & 3 & 0 & & &  \multirow{1}{*}{\}~~~~4} & & \\ 
\bottomrule[1.5pt]
\end{tabular}
}
\caption{The present Table is derived from the Fig \ref{FigXX}. The tracks on the left side and the cascades on the right side are counted in their different zenith angle views. For sake of simplicity we recall here the counting for a couple of more narrow ranges: \SIrange[range-units = brackets]{0}{7.5}{\degree} with 17 upward events, and \SIrange[range-units = brackets]{-7.5}{0}{\degree} with 8 downward.
}
\label{tab:table1}
\end{table}


\section{Conclusions}\label{sec:conclusions}

We were able to conjointly explain the weak or absent overall correlation of the recorded data
with the (catalogs) of radio, X, $\gamma$, GRB source maps, 
the absence of the Galactic Plane signature and of self-clustered neutrino
events, supporting the interpretation of the main part of the events as atmospheric charmed charmed ones.

 An astrophysical origin assumption should lead to a democratic 
 flavor ratio $1:1:1:(1)$ for $\nu_e:\nu_{\mu}:\nu_{\tau}:(NC)$ not remarked and statistically unlikely with the available evidences. Too few tau events both in the data up to 2018, and with the additional data of 2019. We underline that the charmed atmospheric spectra has the same exponent power $-2.7- 2.9$ as Cosmic Rays and as the HESE neutrino signals. This soft spectra may better hide the eventual Glashow resonance signal absence.

 With enough straightforward considerations about the expected symmetry of upward and downward  astrophysical events and the contrariwise data results, notably and not limited to the $\pm \ang{15}$ and the $\pm \ang{7.5}$ angular windows,  we added further indications in favor of the atmospheric origin interpretation.


Moreover, since the noticed relative abundance of upwards vs downwards track events  contrary  to an astrophysical assumption, that we understand as a result of some anticoincidence trigger selection procedure, we are implicitly propounding to look for astrophysical signals whenever possible in the otherwise discarded background data.

The asymmetry among cascades and tracks ratio in upward and downward sky led us to claim a very meaningful small probability  see in Eq.\ref{formula:cas-track}: $5.7 \cdot 10^{-4}$.
The  reasonable explanation for this small amazing ratio is the suppression  made by a veto hiding downward muon tracks. The veto is probing the main atmospheric charmed nature of the downward muon tracks.

Restating the key role of tau in the neutrino astrophysics, we may help improve the complex understanding that underlies the updating and planning of the next-generation detecting systems and to analyze this kind of data.

The archived results will allow, with the availability of a more complete and accurate data set, to refine the goodness calculations for the assumption of the astrophysical vs atmospheric origin of the detected events, as well as to include new decay mechanisms in the procedural research for astrophysical events within the data already collected.

A larger statistical record for well tested tau and for muon tracks may convince the whole scientific arena about the importance of \IceCube  discover, either $\nu$ astronomy or new charmed channels. We believe here the discovering of the charm atmospheric signals. Future more abundant  records may offer more strong source correlation and or the ratio between astrophysical and charmed signal.

%

\section*{Acknowledgements}
The authors would like to thank for the suggestions the Prof. Barbara Mele and Aleandro Nisati for the estimate of the radiative corrections in the Glashow 
diagrams.

This article honors the recent 9th May 2019, both the 74\textsuperscript{th} Victory day for Russia (ex URSS) and  the 71\textsuperscript{st} Independence day for Israel. 

The work by M. Yu. Khlopov was supported by grant of the Russian Science Foundation (project No-18-12-00213).

\bibliographystyle{JHEP}

\bibliography{NoTauNoAstronomy2019-03}


\par\noindent\rule{\textwidth}{0.4pt}

\section*{DISCUSSION} 

\bigskip
\noindent {\bf ULI KATZ:} The events observed  by \IceCube are not consistent with atmospheric prompt neutrinos since these could be accompanied by atmospheric muons that are filtered out by HESE veto.

\bigskip
\noindent {\bf DANIELE FARGION:} It is true that prompt charmed neutrinos might come with their twin muon companion, but only  in less than half of the downward muon neutrinos sky. Horizontal and Upward muon neutrinos are not coming by sure with any muon (absorbed) anyway. Now consider the following question:
 Why highest energetic ($> 90$) TeV downward muon tracks are also almost absent (two events versus 24 showers) while they are well observed (6 tracks versus 10 showers) in  upward sky? The answer coulb be because most of them are not of astrophysical nature, but they are charmed ones and excluded and filtered by  \IceCube veto. Moreover, there is wide agreement in different articles that the atmospheric expected prompt flux is not well defined up to a factor of two. Therefore, prompt neutrino may well explain several puzzles at once.

\bigskip
\noindent {\bf ULI KATZ:} The numbers of good tau neutrinos candidates passing the corresponding event selection is expected to be small, about two events for the full data sample. It is therefore, much too  early to draw conclusions from the fact that no such events are observed so far.

\bigskip
\noindent {\bf DANIELE FARGION:} I agree that the expected two events (since June 2018, possibly observed and reported in Neutrino 2018) within nine events above 200 TeV  might be consistent in a first view with the expected ones. However, the same June 2018 Neutrino report informed us that there are many more (27) events candidate above 100 TeV that are in principle showing a double bang; they are possibly nine tau ones. Their low number (or absence)  is quite surprising.
    More new recent suppression efficiency in tau detection may offer the escape road to the puzzle.  Anyway  the atmospheric charmed tau would be, within 34 showers above 90 TeV energy, anyway about one or two events. Therefore, we believe that the missing tau is still a persistent puzzle, finding a natural solution in a dominant charmed component.

\end{document}